\documentclass[12pt]{article}
\usepackage{graphicx}
\begin{document}
\centerline{\bf Extinction in genetic bit-string model with sexual recombination}

\bigskip
D. Stauffer$^1$ and S. Cebrat$^2$

\bigskip
\noindent
$^1$ Institute for Theoretical Physics, Cologne University

\noindent
D-50923 K\"oln, Euroland

\medskip

\noindent
$^2$ Department of Genomics, Institute of Genetics and Microbiology

\noindent
University of Wroc{\l}aw, ul. Przybyszewskiego 63/77

\noindent
PL-54148 Wroc{\l}aw, 
Poland \\

Abstract:

We have analyzed the relations between the mutational pressure,
recombination and selection pressure in the bit-string model with sexual
reproduction. For specific sets of these parameters we have found three
phase transitions with one phase where populations can survive. In this
phase, recombination enhances the survival probability. Even if recombination
is associated, to some extent, with additional mutations it could be
advantageous to reproduction, indicating that the frequencies of
recombinations and recombination-associated mutations can self-organize in
Nature. Partitioning the diploid genome into pairs of chromosomes
independently assorted during gamete production enables recombinations
between groups of genes without the risk of mutations and is also
advantageous for the strategy of sexual reproduction.

Keywords:

\section{Introduction}

The asexual Eigen quasispecies model \cite{eigen} of biological evolution can be
simulated by identifying each genome with a string of $L$ bits which are either
zero or one. The fittest genome has zero everywhere, and its mutants with some
bits set to one are less
fit. Usually it has been treated in the approximation of a constant population
but extinction studies were also made \cite{malarz}. With every mutation 
decreasing the survival rate by a factor $x < 1$, this factor $x$ was adjusted 
in \cite{pmco} such that the population stayed constant. In both the latter
model \cite{book} and the original Eigen model, a transition was observed 
for increasing mutation rates between an average genome with few mutations,
and one with the number of mutations being large and proportional to $L$. 
This transition is sharp only for $L \rightarrow \infty$.

In the present work, instead we keep the survival factor $x$ fixed and allow the
population to fluctuate. Thus we investigate the mutational meltdown \cite{lg}:
Can the population survive, or do the mutations cause its extinction? This 
transition is sharp only for population size going to infinity, instead of 
the length $L$ going to infinity for a sharp Eigen runaway (error catastrophe).

Moreover, we simulate sexual reproduction with half of the population male and 
the other half female, and each individual having two bit-strings of length $L$
each. We search for the optimal recombination rate, both when recombination
between these two bit-strings does lead to additional mutations at the crossover
point, and when it does not.

We first define the model, then present its numerical results, then explain 
some of them by a simple theory, and finally we summarize our results.

\section{Model}

Each genome is represented by two bit-strings of length $L$ = 8, 16, 32, or 64;
each bit can be zero or one such that zero is a healthy gene and  one represents
a detrimental irreversible mutation. Instead of a single gene a bit could also 
represents several connected genes, or a larger part of a gene, such that the
probability of a back mutation from one to zero is negligible. Thus if a mutation
hits a bit set already to one, this bit stays at one. All mutations
are recessive, and thus only a pair of corresponding one bits on the same locus 
reduces the survival probability per iteration by a fixed factor $x$. 
Survivors have a fixed number $B = 1$ or 4 of offspring per iteration. 
The model keeps the population size $N(t)$ from diverging by a Verhulst death 
probability $V = N(t)/K$ due to a finite carrying capacity $K$. 

Half of the population is male, the other is 
female. Mutations happen at birth with a probability $M$ per bit-string, and
affect from then on child and parent. 
After these mutations, recombination happens with probability $R$ such that the 
first $y$ bits of one bit-string are combined with the last $L-y$ bits of the 
other bit-string, and also the remaining parts are recombined. Now, with 
probability $M_R$, at one side of the crossover point an additional mutation 
happens in both bit-strings. Then one of the two recombined bit-strings is 
selected as gamete; the female selects randomly a male, and a gamete from this 
male together with a gamete from this female forms the genome of their child.
Thus, if each bit-string 
got one new mutation, it is possible that after recombination one bit-string
carries both new mutations and the other bit-string carries none of them. 
The Verhulst survival probability $V = 1 -N(t)/K$ is applied twice, 
to the babies by reducing the effective birth rate, and later at each iteration 
to the adults; this second application may correspond to density-dependent 
infections. 

Thus the survival probability per iteration after birth is $V \cdot x^n$ if $n$ 
is the  number of active mutations, i.e. of one-one bit pairs.

The standard sexual Penna program \cite{book}
is first simplified by omitting the ageing 
interpretation, the pregnancy period, the male fidelity, the dominance, and the 
threshold (limit) for the allowed number of bad mutations. Thus we have two 
bit-strings of length $L$, with all positions equivalent, and only
recessive bad mutations. Instead of the sharp limit, the exponential decay of
survival probability $x^n$ is used. Thus the model has a birth rate $B$, a
mutation probability $M$ (per bit-string), a recombination probability $R$, a 
recombination-mutation probability $M_R$ that after each crossover one bit 
adjacent to the crossover position is mutated in each bit-string, and a birth 
rate $B$. The Verhulst parameter $K$ limits the population. 

\section{Results}

\begin{figure}[hbt]
\begin{center}
\includegraphics[angle=-90,scale=0.5]{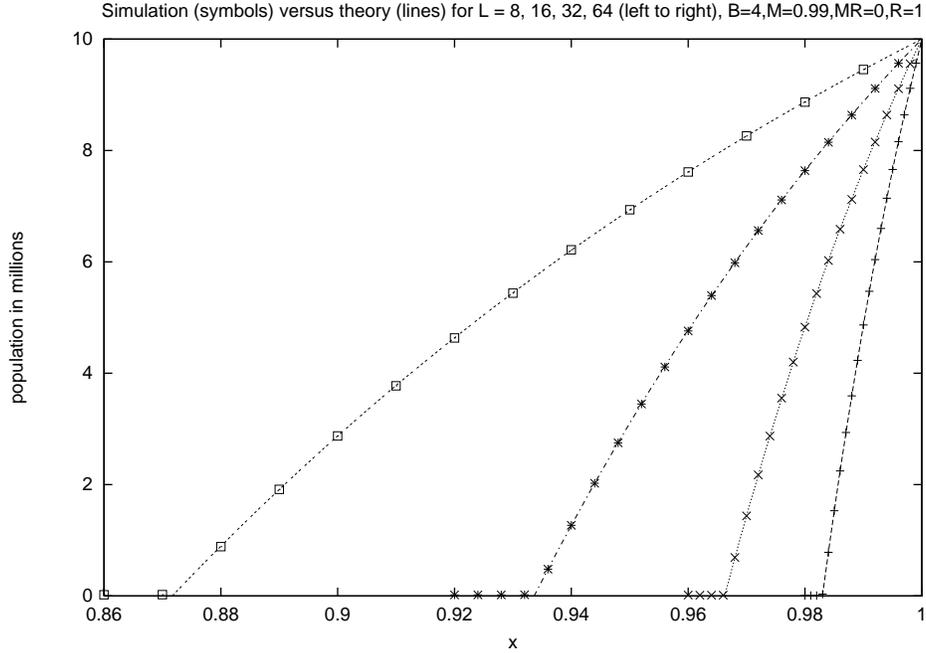}
\end{center}
\caption{Transition between extinction (left) and survival (right), for 
$L=8$, 16, 32 and 64 from left to right. Birth rate $B = 4$, mutation rate 
$M = 0.99$, recombination rate $R=1$, no additional mutations associated with
recombination ($M_R=0$), $K = 20$ million. The curves give 
the theory, Eq.(3).
}
\end{figure}

\begin{figure}[hbt]
\begin{center}
\includegraphics[angle=-90,scale=0.5]{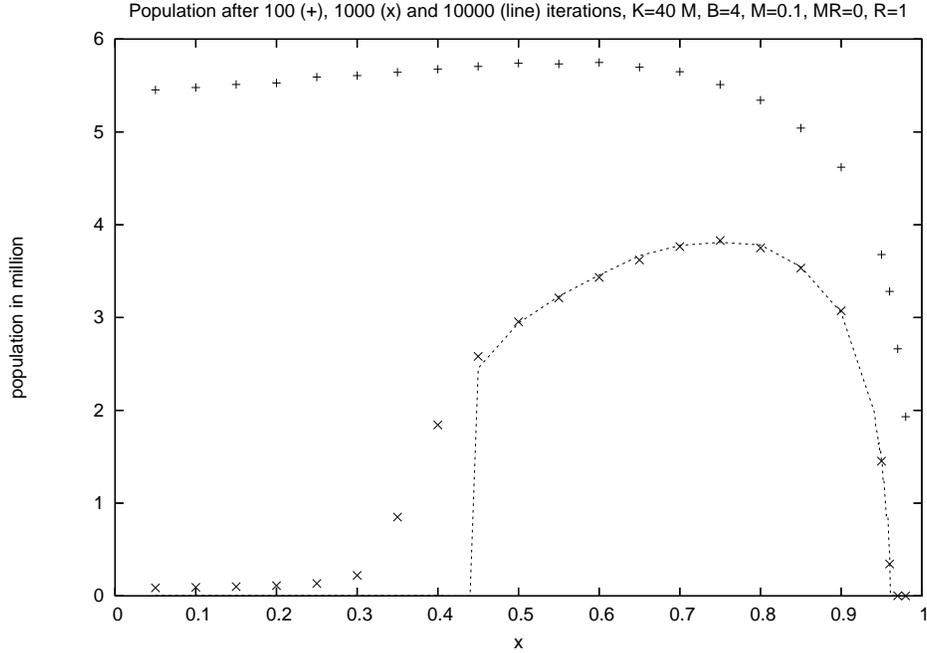}
\end{center}
\caption{Parameters as in Fig.1 except $M=0.1$ instead of 0.99, and $K=40$ 
million, for 
$t = 100$, 1000 and 10000 and $L = 64$. For clarity the large populations for 
$x > 0.983$ are not shown. Curves at $t=10000$ for 
$L = 8$, 16, and 32 look similar in the right part but extend down to $x=0$ on
the left border with roughly constant population.
}
\end{figure}

\begin{figure}[hbt]
\begin{center}
\includegraphics[angle=-90,scale=0.5]{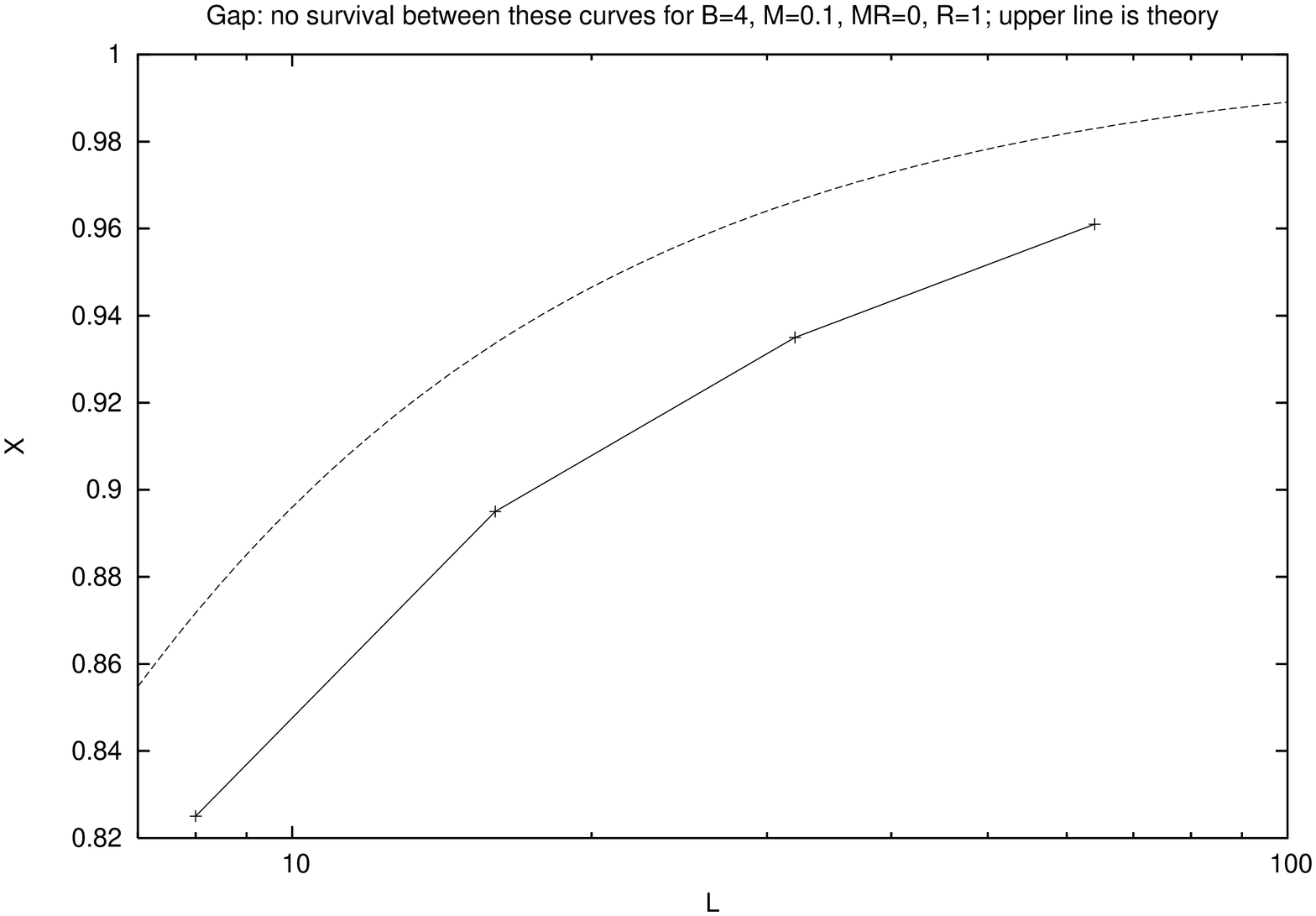}
\end{center}
\caption{
Above the upper curve and below the lower curve, survival is possible
for the parameters of Fig.2. The upper curve is the theory $X = 1/3^{1/L}$ and
agrees with the simulations. 
}
\end{figure}

For $x$ close to one we found that the populations survive and have all bits
set after $ t \sim 10^2$ time steps. For smaller $x$ the populations die out,
$N \propto 1/t$, if we wait long enough. (If the new mutations at birth are 
stored only in the child and not in the parent, survival is also possible for
$x < 0.8$.) Mathematically the limits of $t 
\rightarrow \infty$  and $N \rightarrow \infty$ cannot be interchanged; in 
practice we regard the population as extinct if it decays in this way even
if our actual simulations did not reach $N = 0$. 

\begin{figure}[hbt]
\begin{center}
\includegraphics[angle=-90,scale=0.3]{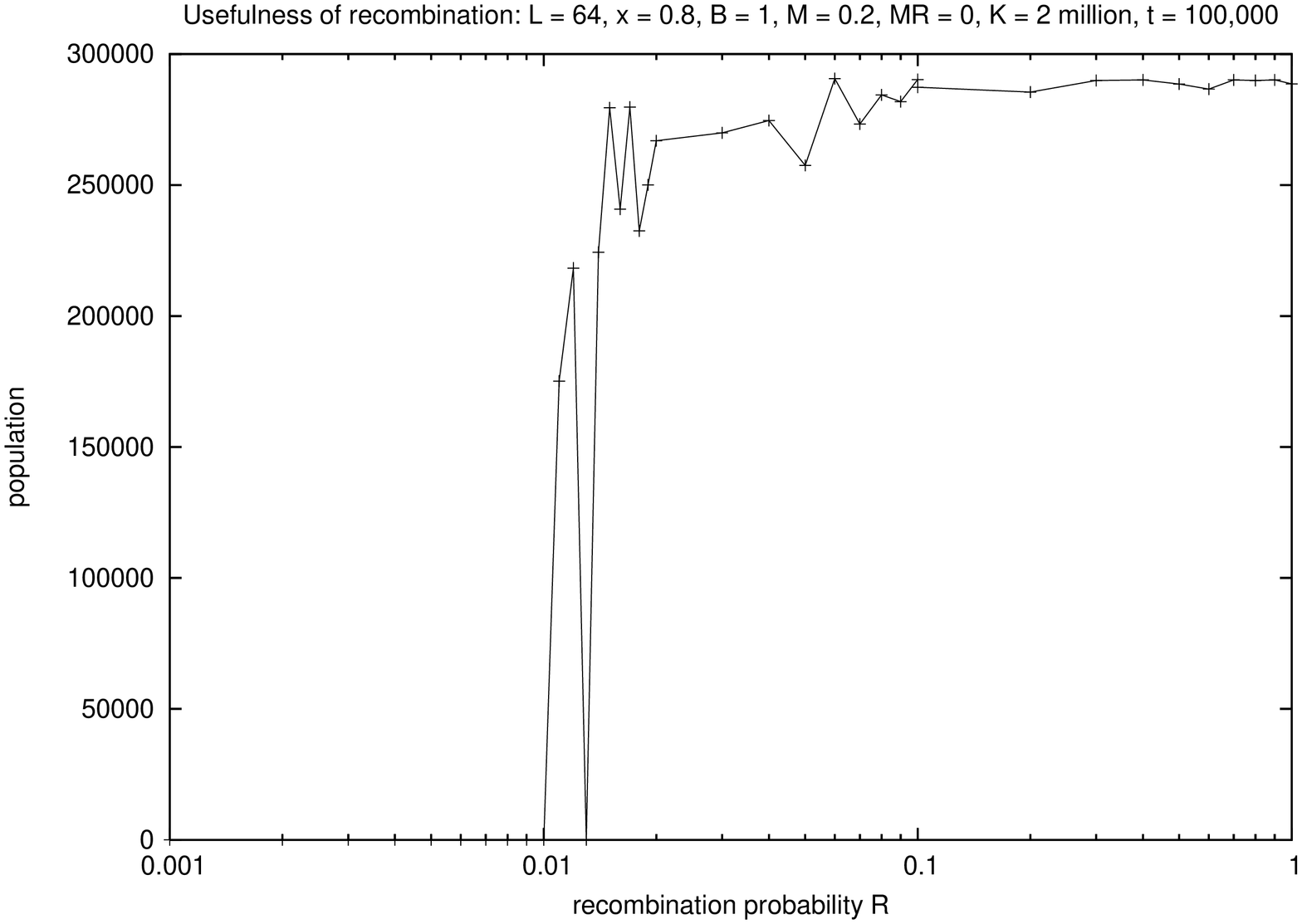}
\includegraphics[angle=-90,scale=0.3]{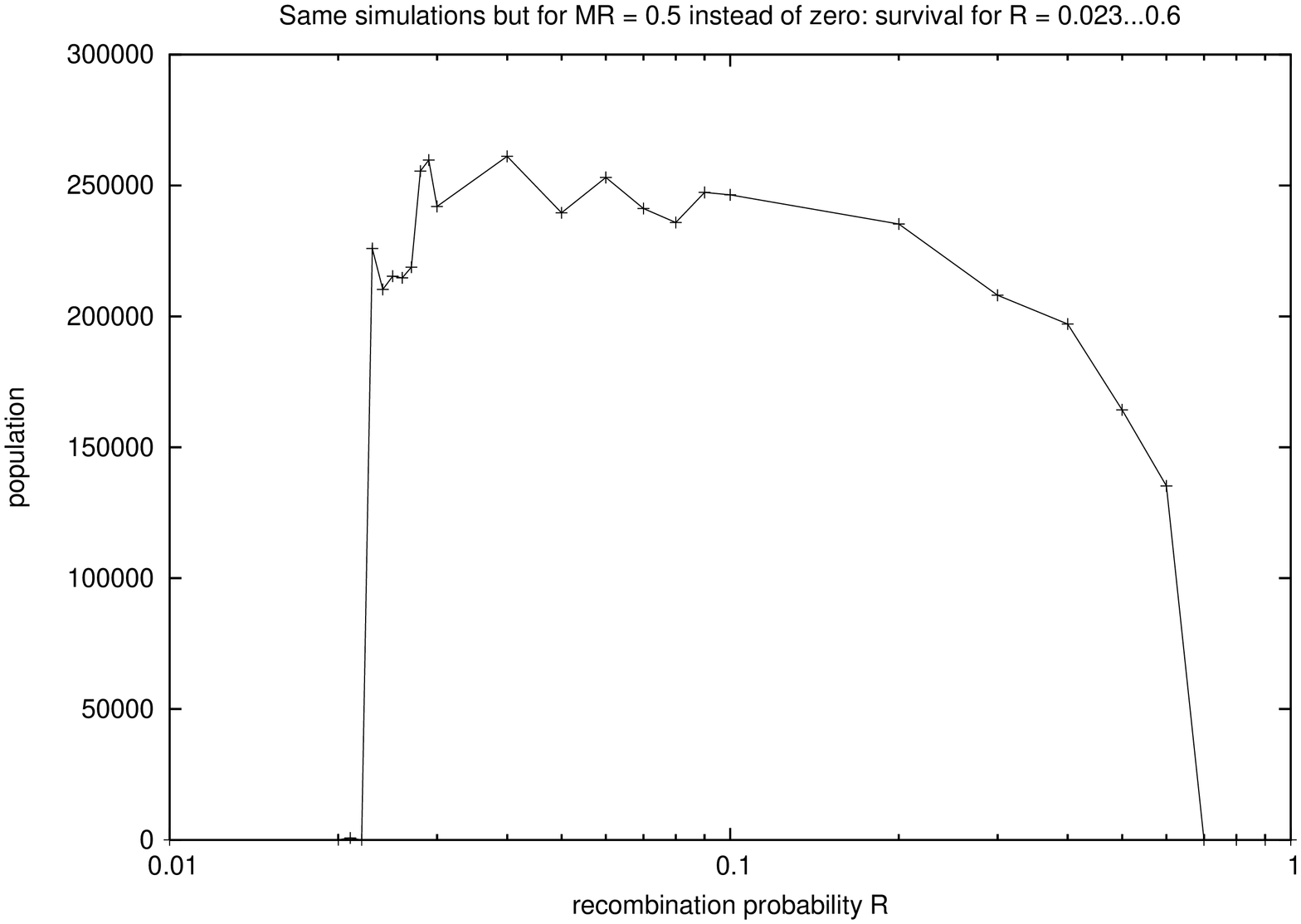}
\end{center}
\caption{The upper part a has $M_R$ = 0, the lower part b has $M_R=1/2$. The 
lower part shows a disadvantage of too small and too large recombination rates.
}
\end{figure}

\begin{figure}[hbt]
\begin{center}
\includegraphics[angle=-90,scale=0.3]{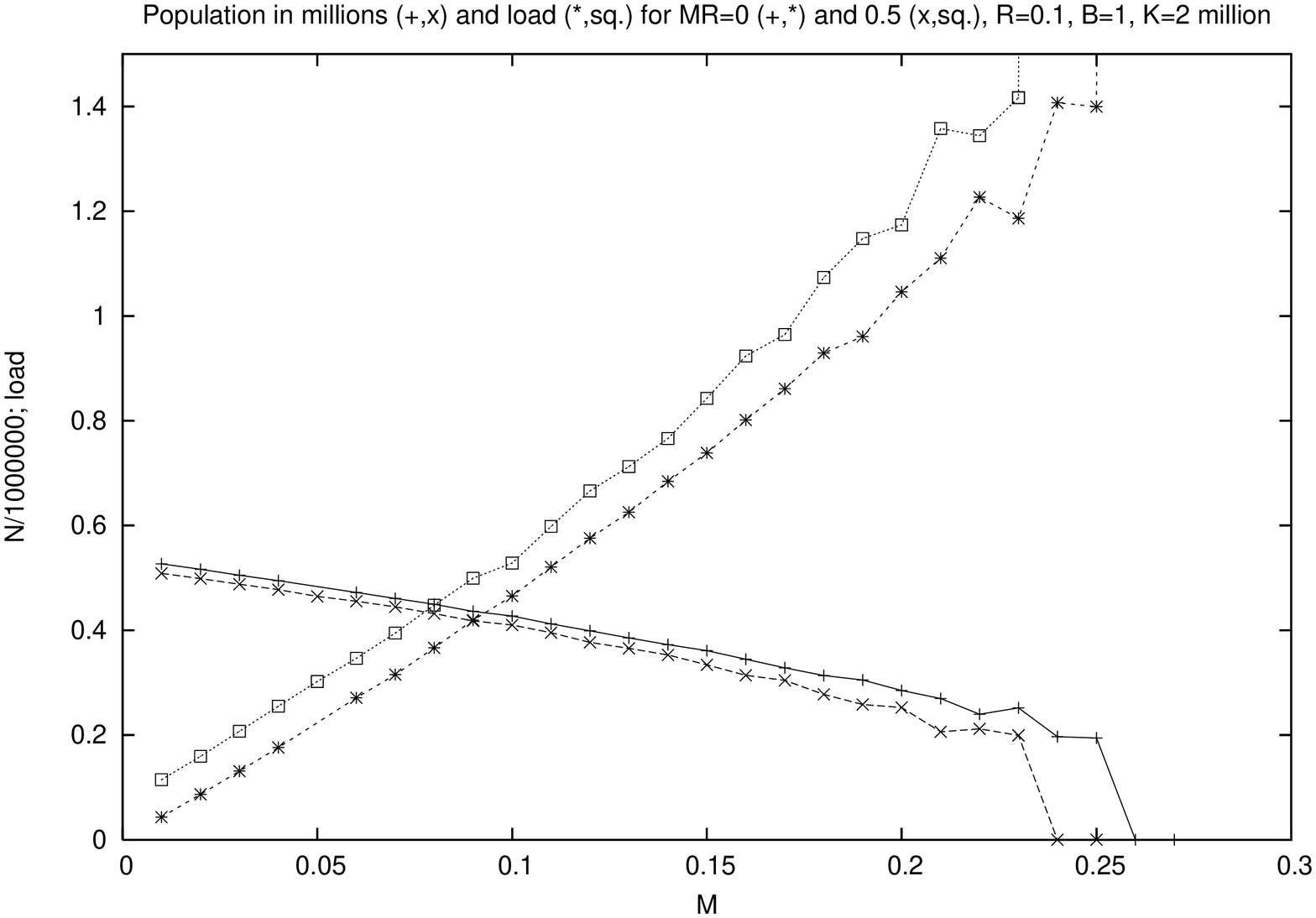}
\includegraphics[angle=-90,scale=0.3]{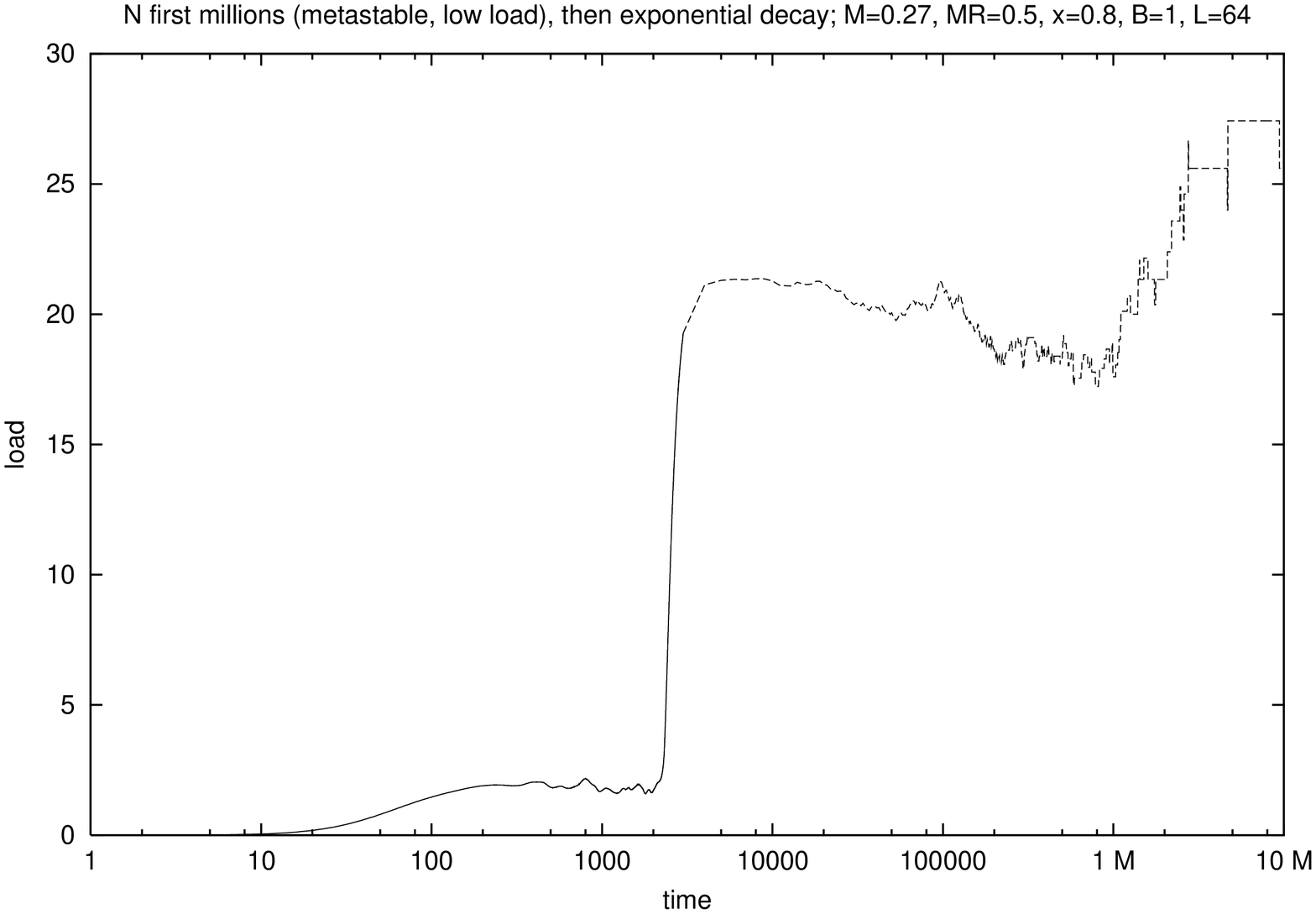}
\end{center}
\caption{Part a: Population and average genetic load versus mutation rate $M$ 
for $L = 64, \; B=1, \; x = 0.8, \; M_R= 0$ and = 1/2. Part b: Dynamics for 
the special case $M = 0.27, \; M_R = 1/2$.
}
\end{figure}

Fig.1 shows the population after 1000 iterations for a large $K$
of 20 million (for males and females together). The lines through the data 
correspond to the theory presented in the next section. Survival for large
$x$ and extinction for smaller $x$ are separated by a sharp phase transition 
for both small and large lengths $L = 8, 16, 32, 64$ of the bit-strings. On the
survival side of the phase transitions, all bits become set in both bit-strings.
On this latter side of high $x$, Darwinian selection of the fittest no longer 
works, and thus this phase near $x=1$ may correpond to paradise more than to
reality.

For the high mutation rates $M \simeq 1$ used in Fig.1, the population dies out 
for all $x$ not close to 1. Using $M=0.1$ instead, also at low $x$ survival 
becomes possible, Fig.2, but the approach to a stationary state is slow. 
The survival regions of small $x$ and of $x$ near unity are separated by 
a small gap, Fig.3. For $L=64, \; M = 0.1$ the survival region has a lower bound
at $x = 0.45$ while for our smaller $L$ all $x$ down to zero allow a survival
under these conditions. The mutation load, i.e. the number of bits set to one,
is small for small $x$, increases
towards the gap, and is maximal $= L$ on the other side of the gap. For
$L=64, \; B=4, \; R=1, \; M=0.1,\; M_R=0$, the mutation load is about 42 for 
small $x$, seems to jump downward with increasing $x$ to 
about 7 near $x=0.45$, then increases slowly to about 31 near $x=0.96$ where it 
jumps to about 46 and then increases until it reaches $L=64$ at $x=0.983$ and 
stays at this maximal value until $x$=1.

As a function of recombination probability $R$ between 0.001 and 1 we see that
for very small $R$ the population dies out quickly; for $R \sim 0.01$ it decays
after a long metastable survival; and larger $R$ prevent extinction, Fig.4a.
If additional bad mutations are associated with each recombination, by setting
$M_R = 1/2$ instead of zero, then also for $R$ near unity the population dies 
out, Fig.4b.

Thus far we always started with an ideal genome. Starting instead with a random
genome (half of the bits mutated) 
and large $x$, the equilibrium results are the same for survival since again 
all bits of both bit-strings become set to one after long enough time. On the
extinction side of the phase transition, the time dependence of the population
is different at first, but later becomes similar though not identical, to the
case when the initial genomes are ideal.
For smaller $x$ like 0.97 (not shown), the population first decays, then
recovers, and finally dies out, if we start with random bit-strings. Fig.5 shows
at fixed survival factor $x = 0.8$ the variation of the population and of the 
genetic load with the mutation rate $M$ at $M_R = 0$ and = 0.5. Too high $M$ 
kill the population by increasing the genetic load after a metastable state with
low load and high population; however, the high load during the decay is 
appreciably smaller than $L$.

Finally, we simulated two instead of only one chromosome by a simple 
approximation: The total length $L = 64$ of the bit-strings is divided into two 
parts, one with $L_1$ bits and the other with $L_2=L-L_1$ bits. Before each 
random crossover also a deterministic pseudo-crossover 
at the bit position $L_1$ is made (always or with probability 1/2); this
deterministic pseudo-crossover is not accompanied by additional mutations there.

The resulting Fig. 6 shows the populations under the same conditions as 
in Fig.4b. Again, too high recombination rates are bad; at low recombination 
rates survival is possible (except $L_1=L/4$ and very low $R$ near 0.001) 
while before the population became extinct for recombination rates of 0.02 and
below. It does not matter much whether the division is 32 + 32 or 16 + 48,
or whether the additional "crossover" is made always or with probability 
1/2.

More numerical results are available in a report from stauffer@thp.uni-koeln.de.

\begin{figure}[hbt]
\begin{center}
\includegraphics[angle=-90,scale=0.5]{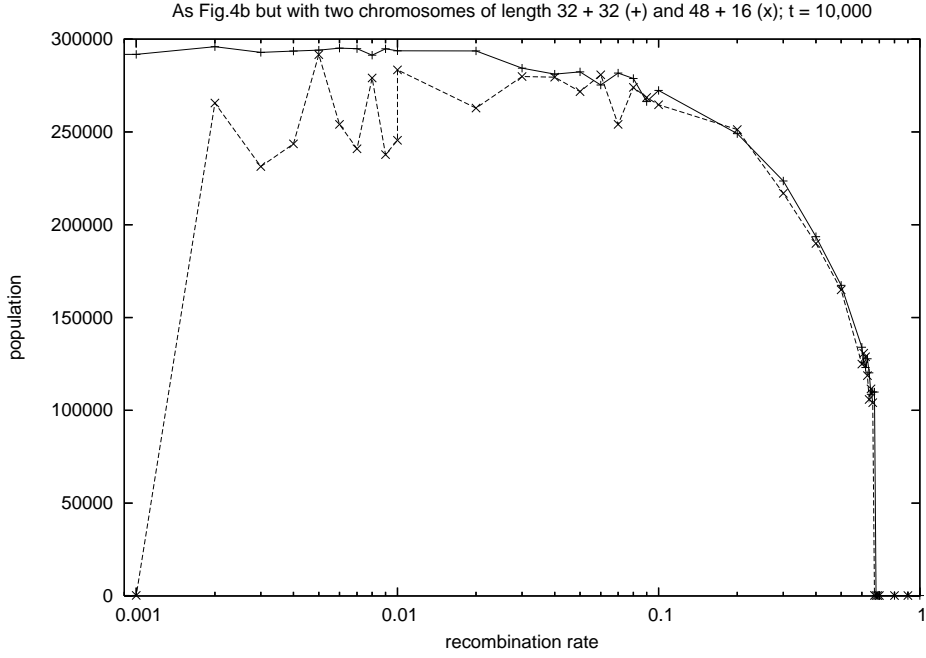}
\end{center}
\caption{
Parameters as in Fig.4b, but now better survival chances with two chromosomes
of length 32 + 32 (+) or 16 + 48 (x); $t = 10^4$.
}
\end{figure}
\section{Simple scaling theory}

For $x$ close to one all bits were set to one and the population survived 
nevertheless. 
Since in this case all further mutations and all crossovers do not change 
anything, all heridity correlations have vanished and a mean-field probability
approach should be valid. This region and its transition point to the extinction
region for smaller $x$ can be explained by a simple scaling theory.

The rate at which daughters are born is $B/2$ if the total birth rate 
(including sons) is $B$. Each female has $L$ mutations and survives with 
probability $x^L$. If she survives she has $B/2$
daughters. Thus the equilibrium point $X$ where without Verhulst factor the 
deaths and births just balance is given by

\begin{equation}
  (1 + B/2) X^L = 1
\end{equation}
or $X = (1+B/2)^{-1/L} = 0.872$, 0.934, 966, and 0.983 at $B=4$ for $L=8$, 16, 
32, and 64, respectively, in agreement with the curves in Fig.1. For large $L$ we have $1-X \propto 1/L$.

For $X<x<1$ we need the Verhulst survival probabilities $BV = 1-N(t)/K$ to 
stabilize the simulations, where $N(t)$ is the population and $K$ the carrying 
capacity. For time-independent $N$, we now have
 
\begin{equation}
   V (1 + V B/2) x^L = 1
\end{equation}
or 

\begin{equation}
   N/K = 1 - V = 1 - [(1 + 2B/x^L)^{1/2}-1]/4 
\end{equation}
which for varying $x$ is a function of only the scaling variable $z = L \ln(x)$,
even for small $L$. Fig.1 shows that this expression fits the simulations for 
$L = 8$ to 64 without any adjustable parameter; the mutation and recombination
rates do not enter this equilibrium theory.

This theory applies to the transition point from the gap (extinction) to $x$
close to 1 (survival). What happens in the gap? In principle, equilibrium there 
means extinction
of the whole population. In practice, one can simulate a large population
with a survival factor $x$ only slightly below the transition point $X$ 
where the decay of the population occurs very slowly. Then a long time interval 
can be found where the population diminishes but the average properties
of the survivors are time-independent. There is an
average load slightly below 64, but the closer we are to the phase transition 
at $x = 0.983$ the closer the load is to 64. Thus at least approximately, the 
transition point $X$ from extinction to survival at $x \simeq 1$ agrees with 
the "runaway" point where the mutation load first reaches its maximum $L$, i.e. 
where all bits are mutated. 
Note: runaway means survival, no runaway means extinction near this phase 
transition; for lower $x$, runaway meant extinction. For $x$ 
below the gap, the mutational load no longer reaches $L$ and thus the above 
theory no longer is valid.

\section{Conclusion}

For bit-string length $L = 64$ we found for increasing survival factor $x$ 
four different phases for $B=4, M_R=0, R=1$ in Figs.1 and 2: 

a) the population dies out for $0 < x < 0.45$; 

b) the population survives for $0.45 \le x < 0.96$;

c) the population dies out for $x \le 0.96 < 0.983$; 

d) the population survives for $0.983 \le x \le 1$. 

\noindent 
At the phase transitions between a and b and
between b and c, the population seems to jump, while at the transition between
c and d it is continuous. Phase d and its transition point to phase c can be
calculated mathematically and obey scaling for arbitrary length $L$ through
the variable $x^L$. Phase b is the biologically relevant one. 

In that intermediate survival phase b, recombination is always useful if it
brings with it no further mutations, Fig.4a; but with about one additional 
mutation per recombination event and per bit-string pair, an intermediate 
value of recombination probability like 0.04 is best, and both too high and
too low recombination probability leads to extinction, Fig.4b. In this latter
case, with two instead of one chromosome, at fixed total length $L$, also lower 
recombination probabilities are allowed, Fig.6.

We thank P.M.C. de Oliveira, S. Moss de Oliveira and A. P\c{e}kalski for 
suggestions and criticism, and the European grants COST P10 and GIACS for
supporting out meeting.

\end{document}